\documentclass[a4paper]{jpsj-suppl}
\usepackage[utf8]{inputenc}
\usepackage{txfonts} 
\usepackage{feynmp} 
\title{Quark mass dependence of H-dibaryon}

\author{Yasuhiro \textsc{Yamaguchi}$^{1,}$
 \thanks{Present address:
 INFN Sezione di Genova, Via
 Dodecaneso 33, 16146 Genova, Italy} 
  and Tetsuo \textsc{Hyodo}$^{1}$}

\inst{$^{1}$ Yukawa Institute for Theoretical Physics, Kyoto University,
Kyoto 606-8502, Japan \\
}

\email{
yasuhiro.yamaguchi@yukawa.kyoto-u.ac.jp
}

\abst{
The H-dibaryon is the exotic multiquark state with baryon number 2 and
strangeness $-2$.
The existence of the deeply bound H-dibaryon is excluded by the 
observation of the double hypernuclei.
However
the recent Lattice QCD simulations have found the bound state below the
$\Lambda\Lambda$ threshold with large quark masses by HALQCD and NPLQCD
collaborations. 
In this talk, the quark mass dependence of the H-dibaryon mass is
discussed using 
the pionless effective field theory (EFT) where a bare H-dibaryon field
is coupled with two-baryon states. 
We determine the parameters in this theory by fitting the
recent Lattice QCD results in the SU(3) limit. 
As a result, we obtain 
the attractive scattering length at the physical point where the
H-dibaryon is unbound.
}

\kword{
Exotic hadron, H-dibaryon, Pionless effective field theory,
Baryon-baryon interactions
}

\usepackage{amsmath}	
\begin{document}
\maketitle

\section{Introduction}

The H-dibaryon investigated by R.~L.~Jaffe~\cite{Jaffe:1976yi} is the
exotic multiquark state which has baryon number 2 and strangeness $-2$.
It was predicted as the flavor-singlet dihyperon state with $J^P=0^+$
and the mass $M_H=2150$ MeV that is 80 MeV below the $\Lambda\Lambda$
threshold.
Recently the H-dibaryon has been studied with the Lattice QCD
simulations 
by the HALQCD and NPLQCD collaborations.
HALQCD found the bound states with the pion mass $m_{\pi}=470$-1170 MeV
in the SU(3) limit~\cite{Inoue:2010es,Inoue:2011ai} and resonances with
the the SU(3)
breaking~\cite{Sasaki:2013zwa}.
NPLQCD found the bound states with $m_{\pi}=390$ MeV in 
the SU(3) breaking case~\cite{Beane:2010hg,Beane:2011iw}.
These Lattice QCD simulations show that the H-dibaryon bounds at
large pion mass regions.

The H-dibaryon has been also studied by the
experimental approaches.
Observation of the double hypernuclei $^{\,\,\,\,6}_{\Lambda\Lambda}$He in the
NAGARA event constrains the binding energy of the H-dibaryon as 
$B_H < 6.93$ MeV because the decay process
$^{\,\,\,\,6}_{\Lambda\Lambda}\text{He}
\rightarrow ^4\text{He}+H$ was not
found~\cite{Takahashi:2001nm,Ahn:2013poa}.
This constraint excludes a deeply bound state.
Belle collaboration investigated the H-dibaryon in the
$\Upsilon(1S)$ and $\Upsilon(2S)$ decays and found no peak structure
near the $\Lambda\Lambda$ threshold~\cite{Kim:2013vym}.
STAR collaboration studying the $\Lambda\Lambda$ correlation in the
heavy ion collision obtained the attractive scattering length of 
$\Lambda\Lambda$ which implies no bound state of
$\Lambda\Lambda$~\cite{Adamczyk:2014vca,Morita:2014kza}.
Hence the current experimental analysis shows absence of the H-dibaryon
so far.

The difference between the Lattice QCD results in the large quark mass
regions and the experiments in the physical point can be connected by
studying the quark mass dependence of the H-dibaryon's properties.
In the literature, there have been two interesting studies of the quark
mass dependence.
One approach is the chiral extrapolation technique.
In Refs.~\cite{Shanahan:2011su,Shanahan:2013yta}, the mass of the bare
H-state was evaluated up to the $m_q^{3/2}$ order 
with the one-loop of the Nambu-Goldstone bosons,
and the H-dibaryon was shown to be unbound at the physical point.
However the couplings to baryon-baryon channels were not considered.
The other study is in the framework of 
the chiral effective field
theory~\cite{Haidenbauer:2011ah,Haidenbauer:2011za,Haidenbauer:2015zqb},
where the baryon-baryon scattering was investigated.
In this study, the H-dibaryon was unbound at the physical point.
However the bare H-dibaryon field considered in the chiral extrapolation
approach was not included here.

In this talk, the quark mass dependence of the H-dibaryon mass is
discussed using the low energy effective field theory (EFT) technique. 
The 
scattering length of the baryon-baryon scattering
with the quark mass dependence is
calculated by the pionless EFT~\cite{Kaplan:1996nv,Braaten:2007nq}, where
two-baryon states are coupled with a bare H-dibaryon field. 
The EFT is applicable to the scattering amplitude of the two-baryon
system near the threshold.
In this study, we focus on the flavor singlet channel.
We determine the parameters in this theory by fitting the
recent Lattice QCD results in the SU(3) limit by the HALQCD.

\section{Formalism}
The pionless EFT is applicable near the two-baryon threshold with 
a large length scale of the bound state.
When the length scale $\ell_B=(2\mu B)^{-1/2} $
is much larger than the interaction range, e.g. the pion wavelength
$\lambda_\pi=1/m_\pi$, 
it is possible to describe 
the baryon-baryon interaction as the contact
term.
Here $\mu$ is the reduced mass of the two baryons,
$B$ is the binding energy and $m_\pi$ is the pion mass.
The length scale $\ell_B$ and the pion wavelength $\lambda_\pi$ in the
Lattice QCD simulations 
are summarized in Table~\ref{table:length}.
In all the case
the ratio of $\ell_B$ and
$\lambda_\pi$ is 
$\lambda_\pi/\ell_B<1$ and therefore
we apply the pionless EFT to describe
the baryon-baryon scattering 
at these quark mass regions.

 \begin{table}[t]
  \caption{Properties of H-dibaryon on the Lattice data given by
  HALQCD~\cite{Inoue:2011ai}  
  and NPLQCD~\cite{Beane:2011iw} collaborations.
  Binding energy $B$, pion mass $m_\pi$, baryon mass $M_\Lambda$,
  length scales $\ell_B$, pion wave lengths $\lambda_\pi$ and
  the ratios $\lambda_\pi/\ell_B$  
  are shown.
  $M_\Lambda$ corresponds to the degenerate baryon mass for the SU(3)
  limit and the $\Lambda$ baryon mass for the SU(3) breaking.}
  \label{table:length}
  \begin{center}   
  \begin{tabular}{llccc|ccc}
   \hline
   &Data &$B$ [MeV]& $m_{\pi}$ [MeV]& $M_{\Lambda}$ [MeV]& $\ell_B$
   [fm]&$\lambda_\pi$ [fm]&$\lambda_\pi/\ell_B$ \\ \hline
   SU(3) limit~\cite{Inoue:2011ai} &HAL-1 &37.8& 836.5& 1749& 0.77& 0.24& {0.31}
		   \\   
   &HAL-2 &33.6& 672.3& 1484& 0.88& 0.29& {0.33} \\   
   &HAL-3 &26.0& 468.6&1161& 1.14& 0.42& {0.37} \\   
   SU(3) breaking~\cite{Beane:2011iw} &NPL &13.2 & 390&1170& 1.55& 0.51& {0.33} \\   
   \hline
  \end{tabular}
  \end{center}
 \end{table}

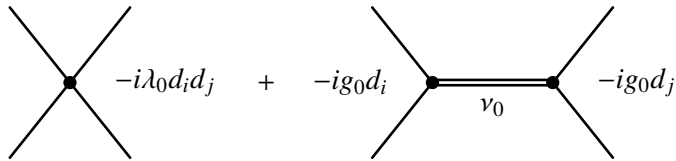
\begin{figure}[b] 
\begin{center} 
\begin{fmffile}{cross_diag_coup_c}
 \unitlength=1mm
 \begin{fmfgraph*}(20,20)
  \fmfleft{l1,l2}
  \fmfright{r1,r2}
  \fmf{plain}{l1,v1,r2}
  \fmf{plain}{l2,v1,r1}
  \fmflabel{\quad$-i{\lambda_0}d_id_j$}{v1}
  \fmfdot{v1}
 \end{fmfgraph*}
\end{fmffile}
 \hspace{1.2cm} \raisebox{0.9cm}{$+$} \hspace{0.5cm} 
\begin{fmffile}{bareH_coup_c}
 \unitlength=1mm
 \begin{fmfgraph*}(40,20)
  \fmfleft{l1,l2}
  \fmfright{r1,r2}
  \fmf{plain}{l1,v1,l2}
  \fmf{plain}{r1,v2,r2}
  \fmf{double,label=${\nu_0}$}{v1,v2}
  \fmflabel{$-i{g_0}d_i$\quad}{v1}
  \fmflabel{\quad$-i{g_0}d_j$}{v2}
    \fmfdot{v1}
    \fmfdot{v2}
 \end{fmfgraph*}
\end{fmffile}
\end{center} 
 \caption{Tree-level diagrams of the four baryons contact term and the
 bare H-dibaryon field.
 }
 \label{fig:tree-level}
\end{figure}

In this study, a four baryons contact term and a bare H-dibaryon field
in flavor singlet channel
are introduced to describe 
the baryon-baryon scattering amplitudes in spin singlet and strangeness 
$S=-2$ sector.
The tree-level Feynman diagrams are shown in
Fig.~\ref{fig:tree-level}.
Here two coupling constants 
appears
in the vertices, $\lambda_0$
and $g_0$ for the contact term and the bare H-dibaryon field,
respectively,
where 
they
are fitted by the Lattice data.
$d_i\, (i=1,2,3=\Lambda\Lambda,\,\,N\Xi,\,\,\Sigma\Sigma)$ is the coefficient for the flavor degeneracy of the two baryon
states in the SU(3) breaking region, 
\begin{align}
 &d_1=\frac{1}{\sqrt{8}},\,d_2=\frac{2}{\sqrt{8}},\,d_3=\sqrt{\frac{3}{8}},
 \quad \sum_{i=1}^{3}d^2_i=1\, .
\end{align}
The primary contribution to the quark mass dependence comes from the
change of the two-body phase space, which is built in the scattering
equation.
Additional quark mass dependence can be introduced in the interaction
parameters.
Here we parameterize the
mass difference $\nu_0$ 
as
\begin{align}
 \nu_0=M^{(0)}_H-2M_\Lambda, \quad M^{(0)}_H =
 M_H-\sigma_H({m^2_\pi}/{2}+m^2_K)\, ,
\end{align}
where $M_\Lambda$ and $m_K$ are the mass of $\Lambda$ and kaon,
respectively~\cite{Shanahan:2011su,Shanahan:2013yta}.
Parameters $M_H$ and $\sigma_H$ are also fitted by the Lattice data.

The scattering amplitude is obtained
by solving the Lippmann-Schwinger equations with 
the contact and bare H-dibaryon field terms.
The diagonal component of the amplitude for the SU(3) breaking is given by
\begin{align}
 & f_{ii}(E)=-\frac{\mu_i}{4\pi}d^2_i\left[\left({\lambda_0}
   +\frac{{g^2_0}}{E-{\nu_0}+i0^+}\right)^{-1}
   +\sum_{\ell=1}^{3}d^2_\ell\frac{\mu_\ell}{\pi^2}\left({\Lambda}-
   \kappa_\ell\tan^{-1}\frac{{\Lambda}}{\kappa_\ell}
   \right)\right]^{-1} \, . \label{eq:amplitude}
\end{align}
$\mu_{i}$ is the reduced mass of two-baryon states for $i=1,2,3$.
$\kappa_\ell$ is given by
 $\kappa_\ell=\sqrt{-2\mu_\ell(E-\Delta_\ell)}$,
where $\Delta_1=0$, $\Delta_2=M_N+M_\Xi-2M_\Lambda$ and
$\Delta_3=2M_\Sigma-2M_\Lambda$.
$M_N$ and $M_\Xi$ are the mass of nucleon and $\Xi$, respectively.
In this study, the SU(3) symmetric interaction is used.
However the SU(3) breaking effect is given by the mass differences of
the baryons.
We introduce the momentum cutoff $\Lambda=300$ MeV in the loop integral.
The flavor singlet scattering amplitude for the SU(3) limit is obtained
by $\Sigma^3_{i=1}f_{ii}(E)$.
The binding energy is obtained as poles of the scattering amplitudes.

\section{Numerical results}

In this study, let us focus on the results which is given by using only
the SU(3) limit data
by HALQCD~\cite{Inoue:2011ai}.
From the Lattice data of binding energies in Table~\ref{table:length}, 
the parameters are 
determined as $\lambda_0=-1.3\times 10^{-5}$
MeV$^{-2}$, $g_0=2.4$ MeV$^{-1}$, $M_H=19783$ MeV and
$\sigma_H=-1.5\times 10^{-3}$ MeV$^{-1}$, respectively.
We obtain the negative $\lambda_0$ which indicates
that the contact term in Fig.~\ref{fig:tree-level} works attractively.
In addition 
the bare mass $M_H$ is large.
The large $M_H$ suppresses the bare dibaryon field term of the scattering amplitude
in Eq.~\eqref{eq:amplitude} and therefore this term plays a minor role
in this system.

From the scattering amplitudes with these fitting parameters,
we calculate 
the quark mass dependence of the scattering length.
In Table~\ref{table:scatleng},
the obtained scattering lengths at the point corresponding to the recent
Lattice data and the physical point are summarized.
We note that the scattering length is the value of 
the baryon-baryon singlet channel 
for the SU(3) limit and of 
the $\Lambda\Lambda$ channel for 
the SU(3) breaking.
For the SU(3) limit,
we obtain the values being close to the
Lattice result~\cite{Inoue:Praivate}.
This indicates that the quark mass dependence of the H-dibaryon and the
low-energy baryon-baryon interaction can be well modeled by the present
EFT.
We also attempt to extrapolate the EFT to the physical point.
We obtain the attractive scattering length at the physical point.
Hence the H-dibaryon is not bound.
A large magnitude of the scattering length supports the validity of the
extrapolation using the contact interaction model.

 \begin{table}[h]
  \caption{Predicted scattering lengths $a$ at the point corresponding
  to the Lattice data by HALQCD~\cite{Inoue:2011ai}
  and the physical point. 
  $M_{B(\Lambda)}$ is the baryon ($\Lambda$) mass at the point.
  Positive (negative) $a$
  shows the repulsive (attractive) scattering length.}
  \label{table:scatleng}
  \begin{center}   
  \begin{tabular}{llcc}
   \hline
   &Point&&$a$ [fm]\\
   \hline   
   SU(3) limit&HAL-1&($M_B=1749$)&1.40 \\[1mm]
   &HAL-2&($M_B=1484$)&1.49 \\[1mm]
   &HAL-3&($M_B=1161$)&1.71 \\[1mm]
   \hline
   SU(3) breaking
   &Physical &($M_\Lambda$=1116) &{$-3.77$}\\
   \hline
  \end{tabular}
  \end{center}
 \end{table}

\section{Summary}
The quark mass dependence of the H-dibaryon has been studied.
The baryon-baryon scattering was described by the pionless EFT
which was applicable near the thresholds.
The scattering amplitude was obtained by solving the Lippmann-Schwinger
equation with the four baryon contact term and the couplings to the bare
H-dibaryon field.
The coupling constant of the EFT was fitted by the recent Lattice QCD
simulations.
By using the Lattice data for the SU(3) limit, the parameters were fixed
and the scattering length were obtained.
At the physical point we found no bound state.
Our model will be improved by introducing the $8s$ and $27$ components,
and the quark mass dependence of the coupling constants.

\section*{Acknowledgments}
This work was supported in part by JSPS KAKENHI Grants No. 24740152 and
by the Yukawa International Program for Quark-Hadron Sciences (YIPQS).

\end{document}